\documentclass[12pt,onecolumn,pre,nofootinbib,floats,aps,amsmath,amssymb,tightenlines]{revtex4-2}

\usepackage{graphicx}
\usepackage{subfig}
\usepackage{color}
\newcommand{\lr}[1]{\left(#1\right)}
\newcommand{\mean}[1]{\left\langle#1\right\rangle}
\newcommand{\vare}{\varepsilon}

\newcommand{\bel}{\boldsymbol{\ell}}

\newcommand{\be}{\begin{equation}}
\newcommand{\ee}{\end{equation}}

\begin{document}

\title{The dynamical role of vortices in the two-dimensional inverse cascade of turbulence}

\author{
V.J. Valad\~{a}o$^{1}${\footnote{Corresponding author: victor.de.jesus.valadao@uniroma2.it}}, 
R.M. Pereira$^{2}$,
L. Moriconi$^{3}$ and 
G. Boffetta$^{4}$}
\affiliation{$^{1}$Department of Physics and INFN, University of Rome “Tor Vergata”,
Via della Ricerca Scientifica 1, 00133 Rome, Italy.}
\affiliation{$^{2}$Instituto de F\'{i}sica, Universidade Federal Fluminense, 24210-346 Niter\'{o}i, RJ, Brazil.}
\affiliation{$^{3}$Instituto de F\'{i}sica, Universidade Federal do Rio de Janeiro, C.P. 68528, CEP: 21945-970, Rio de Janeiro, RJ, Brazil.}
\affiliation{$^{4}$Dipartimento di Fisica and INFN - Università degli Studi di Torino, Via Pietro Giuria, 1, 10125 Torino TO, Italy.}

\begin{abstract}

We study the inverse energy cascade of two-dimensional turbulence by means of high-resolution numerical simulations forced at intermediate wavenumbers. To clarify the role of vortices generated near the forcing scale, we decompose the vorticity and velocity fields into vortex and background components.
Although the vortex component carries only a small fraction of the total energy, it has a clear effect on the energy spectrum and produces deviations from dimensional Kolmogorov scaling, which is realized by the background flow. We derive a generalized von Karman--Howarth balance for the decomposed fields and obtain a relation involving the third-order correlation between vortex and background velocity increments. The simulations support this prediction, which we interpret as a constant inertial-range flux associated with vortex-background correlations.

\end{abstract}

\maketitle


\section{Introduction}
The two-dimensional Navier-Stokes (NS) equations are among the simplest models for large-scale geophysical flows. Despite the absence of rotation and stratification, which are key ingredients of planetary flows, two-dimensional (2D) NS turbulence displays long-lived vortical structures reminiscent of coherent structures observed in the oceans and atmosphere. In decaying 2D turbulence, vortices merge and generate large-scale coherent structures \cite{mcwilliams1990vortices}; in forced turbulence, they are generated mainly near the forcing scale and may slowly grow in time \cite{burgess2022statistics}. These structures coexist with the central phenomenological feature of 2D turbulence: the energy (square velocity) is transferred to scales larger than the forcing, while the enstrophy (square vorticity) is transferred to smaller scales, following the double cascade phenomenology \cite{boffetta2012two}.

In statistically stationary inverse cascades, usually obtained by forcing at intermediate scales and removing energy at large scales, the energy spectrum is observed to follow approximately the Kolmogorov--Kraichnan scaling predicted for a constant energy flux, with small or negligible intermittency corrections \cite{bernard1999three,boffetta2000inverse,muller2025lack}. In the absence of large-scale friction, the inverse cascade is instead unsteady: the total energy grows in time, the integral scale evolves, and the observation of a clean $k^{-5/3}$ range becomes more delicate. An important open question is therefore how forcing-scale vortices affect the statistics of this unsteady inverse cascade. Previous studies have shown that vortices can spoil Kolmogorov scaling and produce deviations in the energy spectrum even at scales larger than the individual vortex size \cite{fontane2013vortical,burgess2015kraichnan}.

In the present paper, we investigate the role of forcing-scale vortices in the statistics of the inverse energy cascade by means of high-resolution direct numerical simulations (DNS) of the 2D NS equation, complemented by theoretical analysis. The forcing is localized at intermediate scales, sufficiently small to allow the development of an inverse cascade and sufficiently larger than the grid scale to allow the development of a direct enstrophy cascade  (which is not the focus in this work). No additional friction term is added to remove large-scale energy. The simulations therefore describe a quasi-stationary, evolving inverse cascade, and they are stopped before the integral scale reaches the box size and energy piles up into a condensate \cite{chertkov2007dynamics,laurie2014universal,frishman2018turbulence,xu2024fluctuation}.

We decompose the total vorticity field into a vortical contribution and a remaining component, which we call the background vorticity field. We find that vortices affect the inertial-range energy spectrum even though they carry a negligible fraction of the total energy. To understand their role in the inverse cascade, we also decompose the third-order velocity structure function, which is proportional to the energy flux. This structure function is dominated by the background field, but it contains a non-negligible vortex-background interaction term. Remarkably, all relevant contributions display linear scaling in the inertial range.

Our results indicate that forcing-scale vortices modify the equal-time statistics of the cascade without carrying the inverse-cascade flux directly. Instead, their main dynamical signature is the growth of correlations with the background field. This suggests that, once the inverse cascade reaches the largest scale of the flow and energy can no longer be transferred to larger scales, the continued growth of vortex-background correlations may contribute to the formation of the condensate. The same decomposition strategy can be applied to other two-dimensional turbulent systems with coherent structures.

\section{Equation and methods}
We consider the 2D Navier-Stokes equation written for the pseudo-scalar vorticity field 
$\omega={\bf \nabla} \times {\bf u}$
\be
\frac{\partial \omega}{\partial t} + {\bf u} \cdot {\bf \nabla} \omega = \nu \nabla^2 \omega + f \ ,\
\label{eq1}
\ee
where ${\bf u}$ is the velocity field, $\nu$ is the viscosity and $f$ represents an external forcing. We consider a square domain of size $L^2$ with periodic boundary conditions and vanishing average vorticity $\langle \omega \rangle=0$ (brackets indicate average over the domain). In the inviscid and unforced limit, (\ref{eq1}) conserves the total energy 
$E=\langle |{\bf u}|^2 \rangle/2$ and the total enstrophy $Z=\langle \omega^2 \rangle/2$. In the presence of forcing, localized at an intermediate scale $\ell_f$, and dissipation, the standard phenomenology of 2D turbulence predicts a direct cascade of $Z$ on scales $\ell<\ell_f$, which is removed at very small scales by viscosity, and an inverse cascade of $E$ towards larger scales $\ell>\ell_f$ \cite{boffetta2012two}. The inverse cascade produces fluctuations at scales larger than the forcing which are not dissipated by viscosity, producing a time-dependent turbulent flow with a total energy that grows linearly in time. This cascade is {\it quasi-stationary} in the sense that there is no accumulation of energy at intermediate scales $\ell>\ell_f$.

Moving to Fourier space, we consider the energy spectrum $E(k)=\pi k \langle |{\bf u}({\bf k})|^2 \rangle$ and its time evolution
\be
\frac{\partial E(k)}{\partial t}=T(k) + F(k) - \nu k^2 E(k)
\label{eq2}
\ee
where $F(k)$ is the forcing contribution and $T(k)$ represents the rate of energy transfer owing to nonlinear interactions \cite{boffetta2012two}. The integral of this latter quantity defines the flux of energy across wavenumber $k$, i.e., $\Pi(k)=\int_k^{\infty} T(k') dk'$, the key observable of the energy cascade. 

From the NS equation (\ref{eq1}), using the assumptions of spatial homogeneity, isotropy, and quasi-stationarity at intermediate scales, it is possible to derive an exact relation for the third-order longitudinal velocity structure function \cite{bernard1999three,lindborg1999can} (see also Appendix~\ref{app:two_point_balances})
\be
S_3(\ell) \equiv \langle (\delta_\ell u)^3 \rangle = \frac{3}{2} \varepsilon \ell \ ,\
\label{eq3}
\ee
where $\delta_\ell u=({\bf u}({\bf x}+\boldsymbol{\ell})-{\bf u}({\bf x}))\cdot \boldsymbol{\ell}/\ell$ is the longitudinal velocity increment over the scale $\ell$ and $\varepsilon$ is the average energy flux in the cascade.
The relation (\ref{eq3}) is expected to hold in scales between the forcing and the integral scale of the flow, i.e., $\ell_f < \ell < L_I$. Assuming self-similarity in the cascade, i.e., a single scaling exponent $h=1/3$ in the cascade, one can predict other statistical objects such as the energy flux in Fourier space $\Pi(k)=\varepsilon$ and the energy spectrum, which follows Kolmogorov scaling \cite{boffetta2012two}
\be
E(k) = C \varepsilon^{2/3} k^{-5/3} \ ,\
\label{eq4}
\ee
where $C$ is a dimensionless constant not fixed by the dimensional analysis. In the unsteady inverse cascade, the integral scale then grows as $L_I^2 \simeq \varepsilon t^3$. After a time of order of $\tau_L\simeq\vare^{-1/3}L^{2/3}$, where $L$ is the domain size, one expects the accumulation of energy at the largest scale of the flow, i.e., condensate formation, idealized first by Kraichnan \cite{kraichnan1967inertial} and later observed numerically \cite{smith1993bose,smith1994finite} and experimentally \cite{paret1997experimental,paret1998intermittency}.

In general, the inverse cascade in 2D turbulence is characterized by the presence of well-defined vortices with linear dimensions prescribed by the forcing scale $\ell_f$ (see Fig.~\ref{fig1}). To investigate their role in the turbulent cascade, we decompose the vorticity field $\omega$ into a vortex contribution $\omega_v$ and a background field $\omega_b$ such that
\be
\omega({\bf x})=\omega_v({\bf x})+\omega_b({\bf x}) \ .\
\label{eq5}
\ee
Together with (\ref{eq5}), we also decompose the velocity field into the associated components as 
${\bf u}({\bf x})={\bf u}_v({\bf x})+{\bf u}_b({\bf x})$.
To implement this decomposition, we use the swirling strength criterion to identify vortex structures \cite{zhou1999mechanisms}. From the velocity field, we compute the eigenvalue of the velocity gradient $\lambda({\bf x})=\lambda_R({\bf x})+i\lambda_I({\bf x})$ as the solution of 
\be
\det\lr{\partial_iu_j-\lambda\delta_{ij}}=0 \ .\
\label{eq6}
\ee
Vortices are then identified as connected regions where $\lambda$ displays an imaginary part, i.e., where
\be
|\lambda_I|\geq \lambda_c \equiv \alpha\sqrt{\mean{\lambda_I^2}-\mean{\lambda_I}^2} \ ,\
\label{eq7}
\ee
where $\alpha$ is a conveniently defined threshold to remove spurious numerical fluctuations. By definition, the background vorticity $\omega_b$ is calculated at positions $|\lambda_I|<\lambda_c$. As a consequence, the total enstrophy can be decomposed into the sum of vortex and background contributions, $Z=Z_v+Z_b$, where $Z_v=\langle \omega_v^2 \rangle/2$ and $Z_b=\langle \omega_b^2 \rangle/2$. This is not true for the energy, since the velocity correlation between vortex and background regions does not vanish. Indeed,
\be
E=\frac{1}{2} \langle |{\bf u}|^2 \rangle = E_v + E_b + C_{bv}
\label{eq8}
\ee
where $E_v$ and $E_b$ are the vortex and background contributions to the total energy and  $C_{bv}=\langle {\bf u}_v \cdot {\bf u}_b \rangle$ is the vortex-background velocity correlation. 

We perform high-resolution direct numerical simulations (DNS) of equation (\ref{eq1}) starting from a zero-vorticity field using a dealiased pseudo-spectral code at resolution $8192^2$. Forcing is localized in a narrow shell of wavenumbers around $k_f=200$, and it is $\delta$-correlated in time to control the energy input rate $\varepsilon_f\simeq 8.5\times10^{-3}$ in simulation units. A detailed explanation of the numerical scheme and its implementation on GPUs can be found in \cite{valadao2024spectrum}. A fraction of the input is dissipated by the (small) viscosity, and the remaining $\varepsilon<\varepsilon_f$ is transferred to large scales by the inverse cascade. Simulations are stopped before this cascade reaches the scale $L$, more precisely at $t\approx0.367\tau_L$.

At every $10^{-3}\tau_L$, we decompose the vorticity field into the two components in Eq.~(5).
The threshold in Eq.~(7) is set to $\alpha=3.5$ from the[] analysis described in Appendix~A, which uses the one-point statistics of the background field and compares the resulting decomposition with alternative vortex-identification methods.

From the vortex fields, at each time, we identify $N(t)$ individual vortices as the connected regions where $\omega_v({\bf x})\neq0$. For the $i$-th vortex, we compute its area $A_i$, from which we define the equivalent radius $R_i=\sqrt{A_i/\pi}$, and other geometrical properties. All the results shown in the following sections are averaged over $9$ independent realizations of forcing to improve statistics.

\section{Numerical results}\label{sec3}

\subsection{Field decomposition}
\begin{figure}[htbp]
\centering\includegraphics[width=0.99\linewidth]{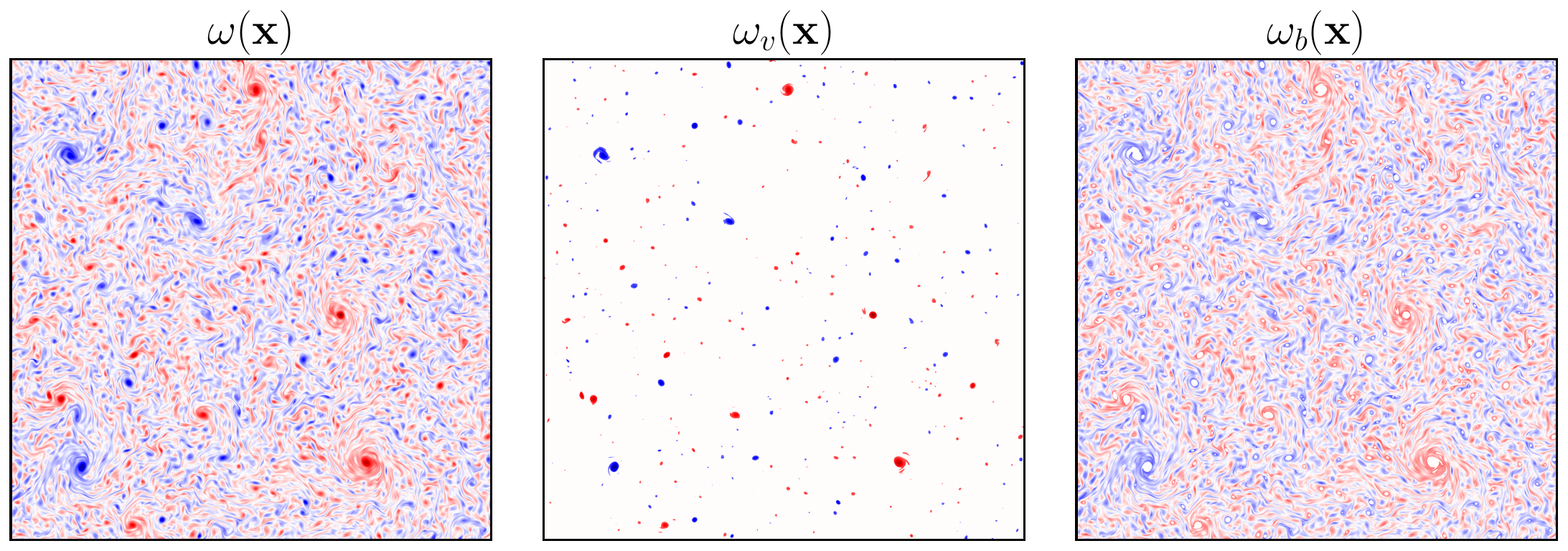}
\caption{Snapshot of the vorticity field at the final time of one simulation, $t/\tau_L \simeq 0.367$. Red (blue) indicates positive (negative) vorticity. Left: total vorticity field $\omega$. Center: extracted vortex contribution $\omega_v$. Right: background contribution $\omega_b$. The panels show a fraction $(L/4)^2$ of the periodic box.}
\label{fig1}
\end{figure}

Figure~\ref{fig1} shows a typical snapshot of the vorticity field and its vortex-background decomposition. The vortex field displays a complex spatial organization, with voids, clusters, and vortex interactions such as merging events. Individual vortices also show large variability in intensity, area, and shape, both in space and in time.

\begin{figure}[htbp]
\centering
\includegraphics[width=0.48\linewidth]{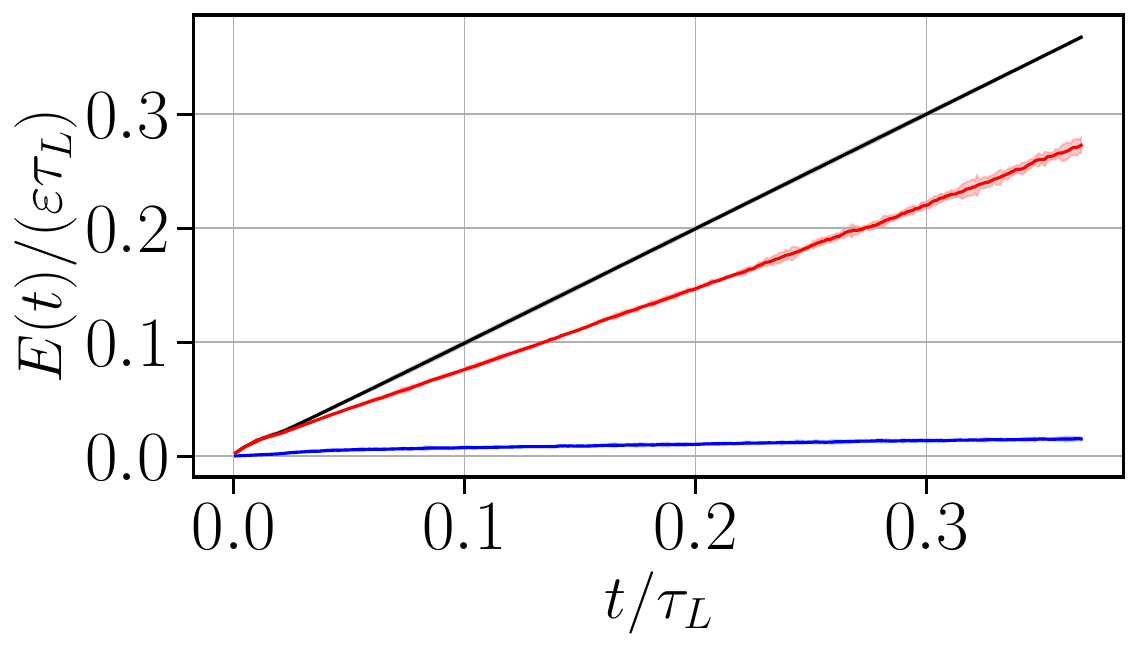}
\includegraphics[width=0.48\linewidth]{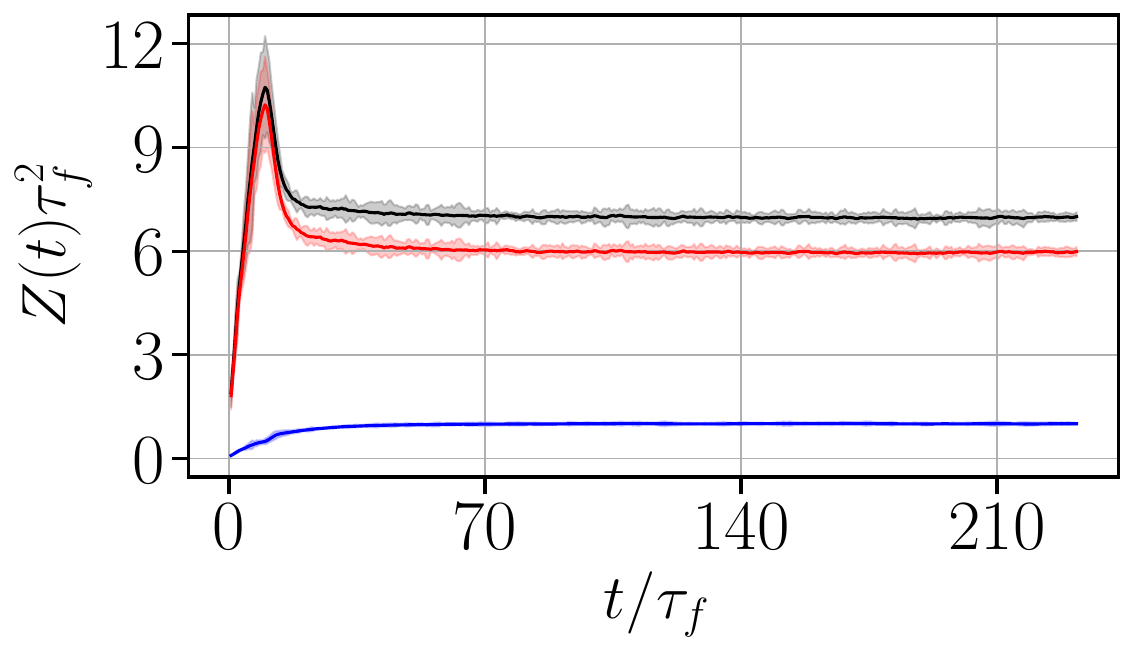}
\caption{Energies (left) and enstrophies (right) as functions of time, normalized with 
$\tau_L=3C/2 \ L^{3/2}\vare^{-1/3}\approx 109$ 
and $\tau_f=(\varepsilon k_f^2)^{-1/3}\approx0.17$, respectively. Black curves show the total-field contribution, while red and blue curves show the background and vortex contributions, respectively. The shaded area shows the spread over the ensemble of 9 realizations.
}
\label{fig2}
\end{figure}

The evolution of the energy and enstrophy, averaged over the different realizations, is shown in Fig.~\ref{fig2} for the two components of the turbulent field $\omega_b$ and $\omega_v$ as well as for the full field $\omega$. The total energy grows linearly in time, as expected for a quasi-stationary inverse cascade. From the slope in the left panel of Fig.~\ref{fig2}, we measure the inverse energy flux to large scales, $\varepsilon=\frac{dE}{dt} \simeq 0.58 \varepsilon_f$. The background field contributes a nearly constant fraction of about $74\%$ of the total energy, i.e., $E_b/E=0.74$, while the vortex contribution is almost negligible, $E_v/E \simeq 0.04$. From Eq.~\eqref{eq8}, the correlation term $\langle {\bf u}_v \cdot {\bf u}_b \rangle$ accounts for about $22\%$ of the total energy.

The right panel of Fig.~\ref{fig2} shows that, for $t/\tau_f \ge 70$, where $\tau_f=(\varepsilon k_f^2)^{-1/3}$ is the typical timescale at the forcing scale, the enstrophy reaches a stationary state in the direct cascade, in which all the enstrophy injected by the forcing is dissipated by viscosity. In this case, the enstrophy is dominated by the background contribution, with $Z_b/Z \simeq 0.86$, while $Z_v/Z \simeq 0.14$ is associated with the vortex field. 

\subsection{Vortex statistics}
\begin{figure}[htbp]
\centering\includegraphics[width=0.46\linewidth]{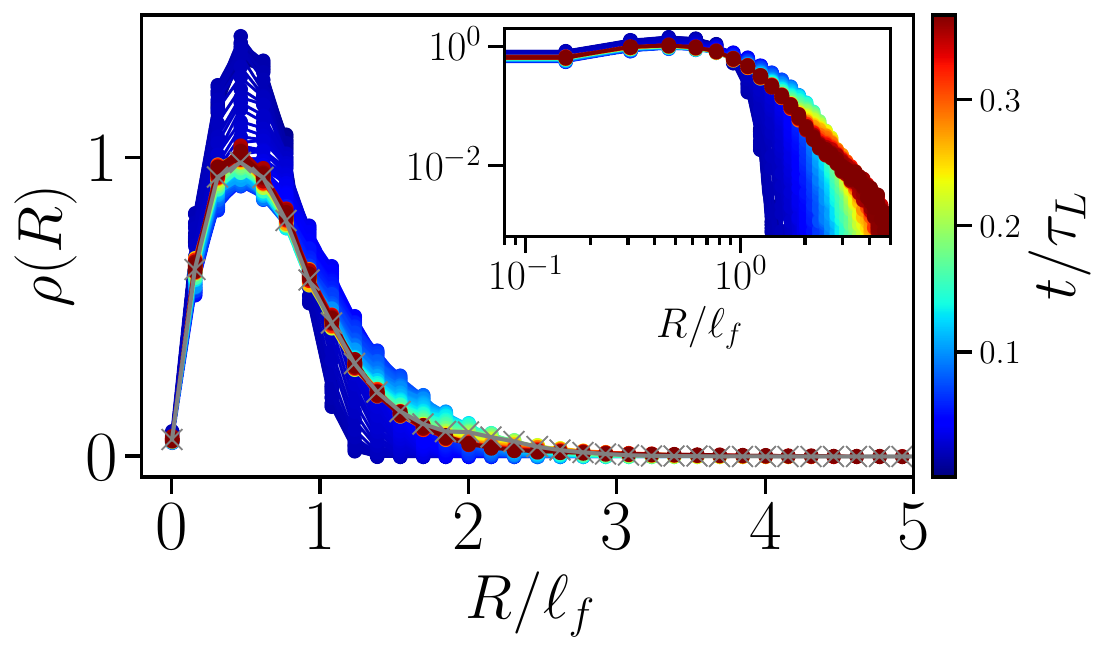}
\centering\includegraphics[width=0.48\linewidth]{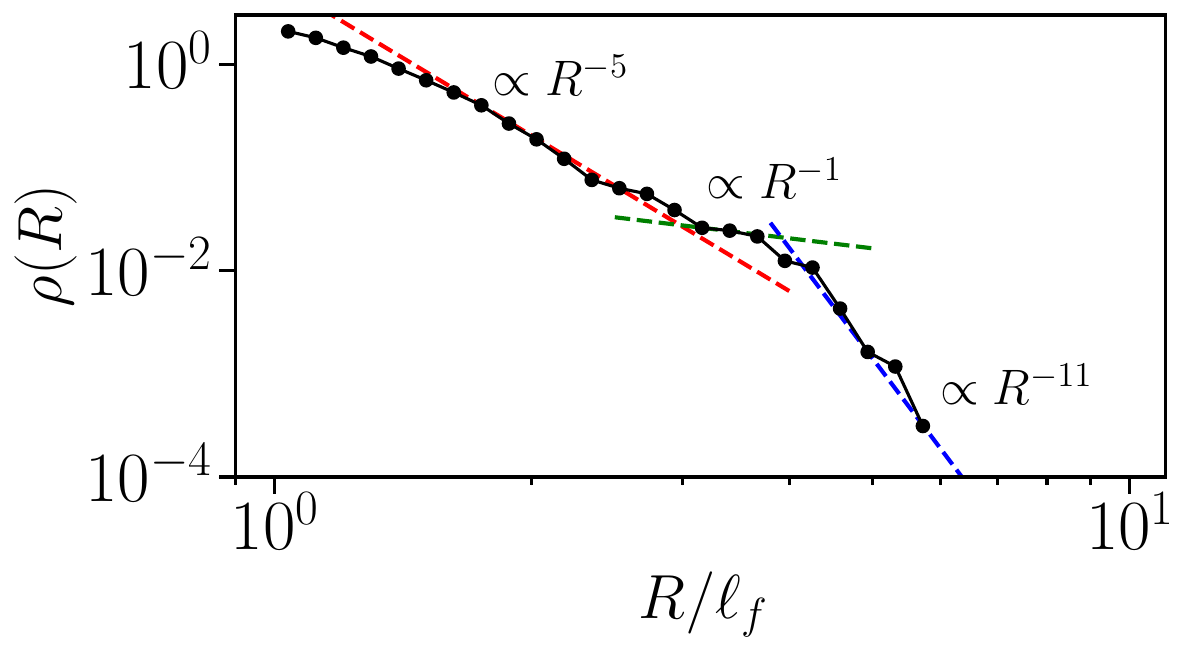}
\caption{Left: PDF of the equivalent radius computed at different times (corresponding to different colors). Gray crosses show the distribution at time $t\approx0.185\tau_L\approx116\tau_f$. The inset shows the same data on a log-log scale. Right: final-time PDF of the equivalent radius. Dashed lines show the predictions from \cite{burgess2017scaling}.}
\label{fig3}
\end{figure}

\begin{figure}[htbp]
\centering\includegraphics[width=0.46\linewidth]{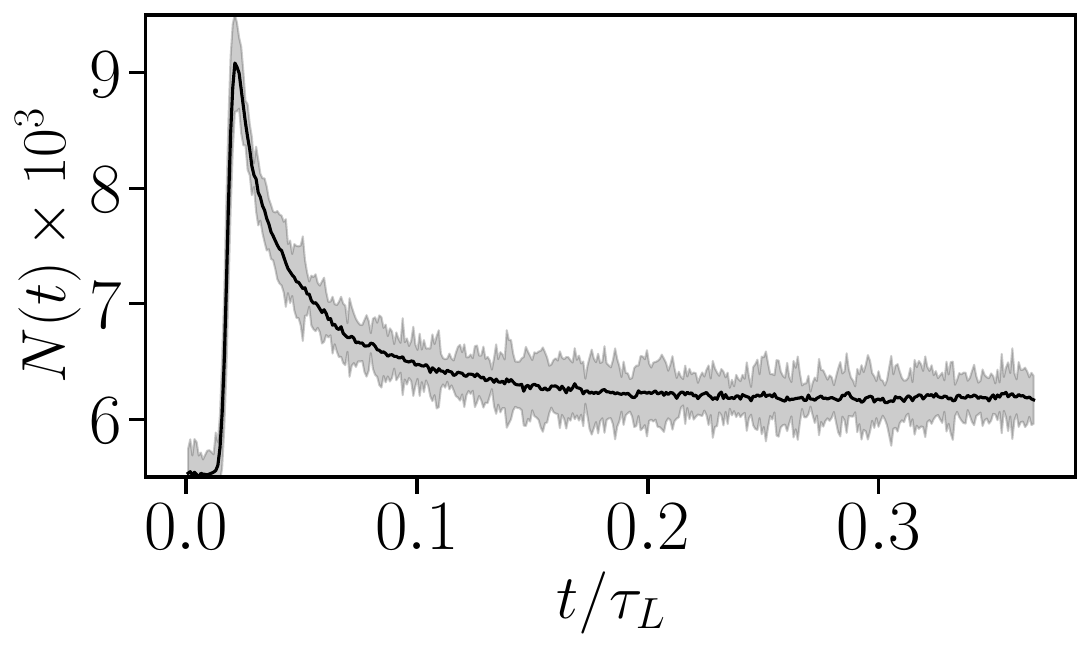}
\centering\includegraphics[width=0.48\linewidth]{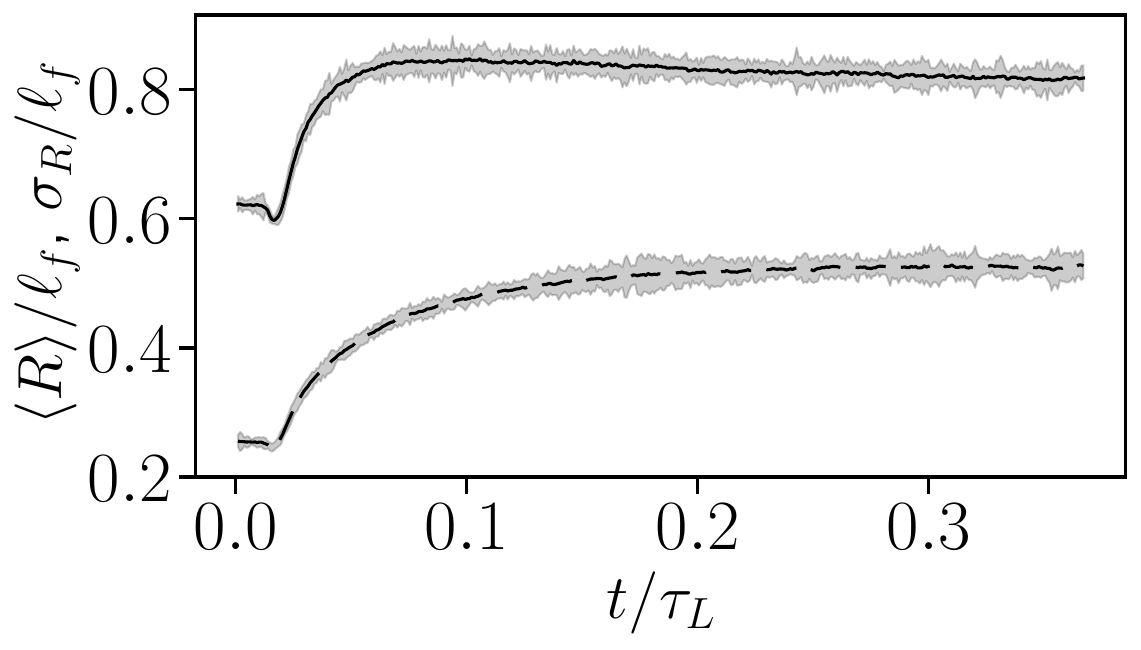}
\caption{Left: total number of detected vortices $N(t)$ as a function of time. Right: the solid and dashed lines yield, respectively, the mean equivalent radius $\mean{R}/\ell_f$ and standard deviation $\sigma_R/\ell_f$ of the vortex-radius PDF. The shaded areas represent the maximum spread over the ensemble of 9 realizations.}
\label{fig4}
\end{figure}

The statistical properties of the individually detected vortices are shown in Figs.~\ref{fig3} and \ref{fig4}. The left panel of Fig.~\ref{fig3} shows the probability density functions (PDFs) of the equivalent radius $\rho(R,t)$ at different times. Two stages are apparent, corresponding to the different enstrophy regimes in the right panel of Fig.~\ref{fig2}. For $t/\tau_L \lesssim 0.1$, the vortex-radius PDF is sharply concentrated around the forcing scale, $R\sim\ell_f/2$. At later times, the distribution remains peaked near the forcing scale but broadens, allowing for larger vortices. This late-time regime corresponds to a fully developed enstrophy cascade, where the core of the single-vortex statistics becomes slowly varying in time. For $t/\tau_L \ge 0.185$ (gray crosses in the left panel), the vortex-radius distribution changes only weakly over the time interval considered.

The right panel of Fig.~\ref{fig3} compares the final-time distribution with the theoretical scaling regimes proposed recently for the vortex population in a forced inverse energy cascade \cite{burgess2017scaling}. This theory describes a population in which vortices generated near the forcing scale interact and merge, progressively populating larger sizes. Immediately above the forcing scale, a forcing-equilibrated ``thermal bath'' approximately conserves vortex self-energy over a scale interval that shifts toward larger areas with the evolving vortex population. This leads to a vortex-area number density $n(A)$ which scales as $n(A)\propto A^{-3}$ at fixed times. At intermediate areas, a scale-invariant vortex-size distribution hypothesis gives $n(A)\propto A^{-1}$ at fixed times. Finally, the largest vortices would form a sparse advancing front described by a steeper distribution $n(A)\propto A^{-r_3}$. The exponent $r_3$, however, is not fixed by scaling arguments and the behavior $r_3\simeq6$ has been numerically observed. At a fixed time, the normalized radius PDF must satisfy $\rho(R)dR\propto n(A)dA$. Since $A=\pi R^2$, these area-space scalings become
\be
\rho(R)\propto R^{-5},\qquad
\rho(R)\propto R^{-1},\qquad
\rho(R)\propto R^{-(2r_3-1)}\simeq R^{-11}.
\label{eq9}
\ee
In our simulations, the final-time PDF is compatible with these radius-space slopes, although over very limited ranges. The comparison should nevertheless be regarded as qualitative, since the temporal scalings and evolving crossover areas predicted by the theory are not tested here. Moreover, both the numerical setup and the vortex-identification procedure differ. In particular, Appendix~\ref{app:threshold_dependency} shows that their identification procedure selects a much sparser vortex population and significantly modifies the measured number and size distributions. Despite these differences, the successive changes in slope suggest that the late-time vortex population contains distinct size ranges qualitatively similar to those predicted.

The total number of detected vortices $N(t)$ is plotted in the left panel of Fig.~\ref{fig4}. After a sharp increase at a short time, corresponding to the increase in enstrophy, this number slowly decreases and reaches an almost constant value for large times $t/\tau_L \gtrsim 0.2$. The first two moments of the radius PDF are shown in the right panel of Fig.~\ref{fig4}. We define
\be
\mean{R}(t)=\int R\rho(R,t)dR,\qquad
\sigma_R^2(t)=\int [R-\mean{R}(t)]^2\rho(R,t)dR .
\label{eq10}
\ee
The mean radius rapidly approaches a nearly stationary value, while $\sigma_R$ grows more slowly over the same interval. As anticipated, the growth of the standard deviation is a quantitative measure of the broadening of the PDF tail. It indicates that rare larger vortices continue to develop even after the number of vortices and the mean radius vary only weakly.

Since one can write the vortex enstrophy
\be
Z_v(t)=\frac{1}{2L^2}\int \omega_v^2(\mathbf{x},t)d^2x
=\frac{1}{2L^2}\sum_{i=1}^{N}\int_{A_i}\omega^2(\mathbf{x},t) d^2x
=\frac{\mean{N}}{L^2}\int \rho(R,t)\mathcal{Z}(R,t)dR \ ,
\label{eq11}
\ee
where 
\be
\mathcal{Z}(R,t)=
\left\langle
\frac{1}{2}\int_{A_i}\omega^2(\mathbf{x},t) d^2x
\right\rangle_{R_i=R}  
\label{eq12}
\ee
is the average enstrophy carried by vortices of radius $R$, the approximate stationarity of $N(t)$, $\mean{R}(t)$, and $Z_v(t)$ suggests that the late-time evolution is not exhausted by changes in the number of vortices or by the typical vortex size. Indeed, the vortex kinetic energy continues to grow,
\be
\frac{dE_v}{dt}>0 \ .
\label{eq13}
\ee
This growth must be connected to the collective organization of the vortex population, such as changes in spatial correlations, clustering, or polarization. This organization is likely associated with vortex-background interactions. In the next subsection, we probe this mechanism by analyzing the spectral distributions of energy and flux in the vortex-background decomposition.

\subsection{Scale-by-scale cascade diagnostics}

\begin{figure}[htbp]
\centering\includegraphics[width=0.48\linewidth]{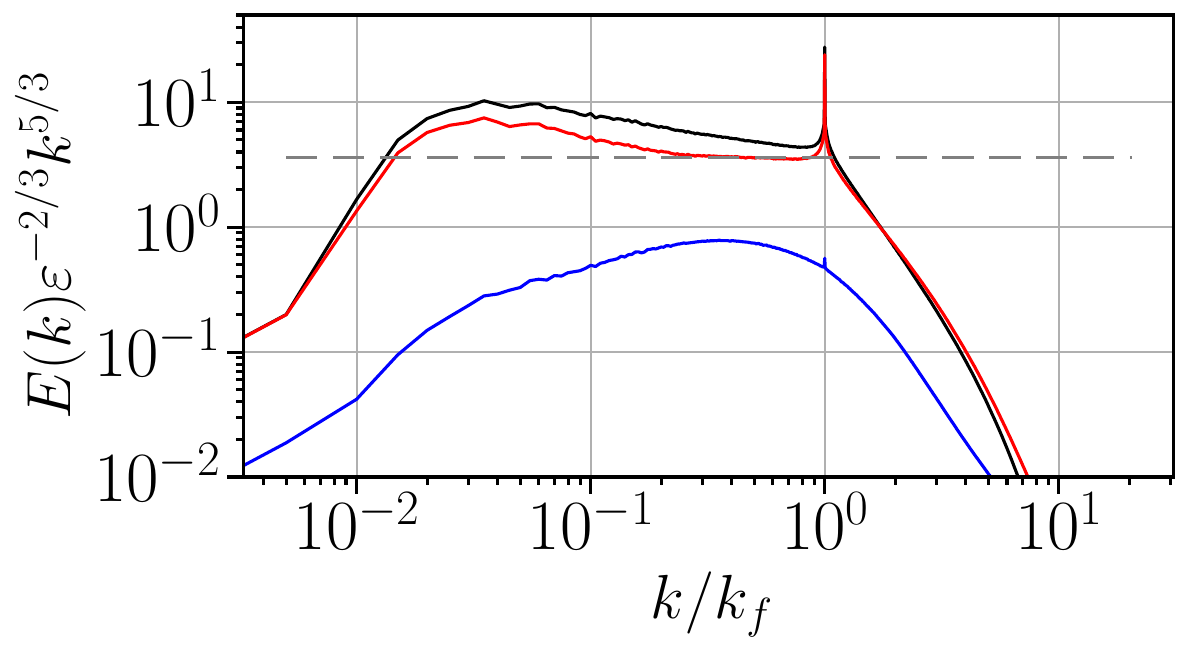}
\centering\includegraphics[width=0.46\linewidth]{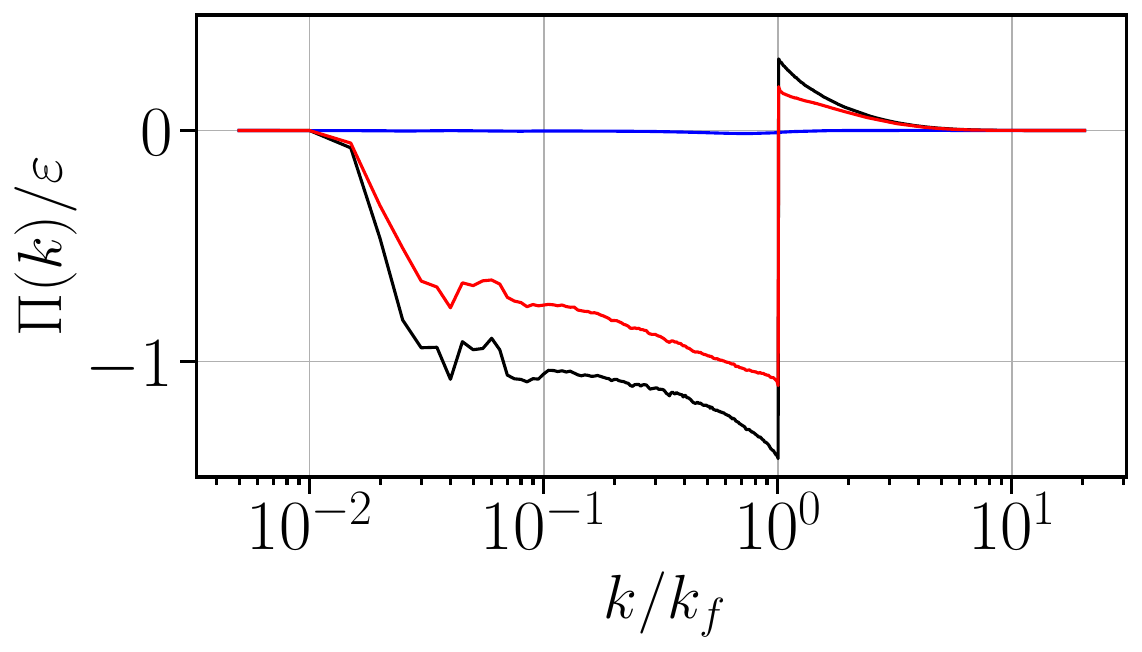}
\caption{Left: compensated spectra; right: nondimensionalized fluxes
computed in the final interval of time $t/\tau_L\in[0.334,0.348]$. The contributions from the total field, background, and vortices are depicted in black, red, and blue, respectively.}
\label{fig5}
\end{figure}

We now discuss how the vortex decomposition affects two-point turbulent diagnostics, both in Fourier and in physical space. Figure~\ref{fig5} shows the compensated total energy spectrum in the final stage of the simulation, together with its decomposition into vortex and background components, compensated by the Kolmogorov scaling in Eq.~\eqref{eq4}. Although the vortex contribution to the total energy is very small, its effect on the spectrum is not negligible. While the total spectrum does not exhibit clear Kolmogorov scaling, the spectrum of the background field $E_b(k)$ fits Eq.~\eqref{eq4} over almost one decade above the forcing wavenumber peak. The effect of small-scale vortices on the scaling of the energy spectrum was already investigated in \cite{fontane2013vortical} and is here confirmed at higher resolution. The left panel of Fig.~\ref{fig5} also shows that the vortex spectrum remains small at all scales. Any approximate scaling visible in this component is therefore not associated with a direct contribution to the inverse-cascade flux, as shown below.

The right panel of Fig.~\ref{fig5} shows the energy flux at the same times, computed from the total field and from the vortex-background decomposition. The vortex field has a virtually vanishing flux, indicating that vortices do not carry the inverse-cascade flux directly. 

The flux associated with the background field, $\Pi_b(k)$, is close to but smaller than the total flux and does not by itself display the full constant-flux plateau. A possible contribution to this feature is the rapid separation of vortex-antivortex pairs after their formation, which may produce the observed energy increase which is observed in the energy flux at scales slightly below the forcing wavenumber.

The comparison between the two panels of Fig.~\ref{fig5} therefore shows that vortices modify the spectral scaling while contributing only indirectly to the spectral energy transfer. The missing contribution to the total flux comes from correlations between vortex and background fields.

The spectral analysis is complemented by structure functions (SFs) in physical space, based on velocity increments. For clarity we introduce a bookkeeping parameter $\epsilon$ in the velocity decomposition and write
\be
\mathbf u = \mathbf u_b + \epsilon \mathbf u_v
\label{eq14}
\ee
By substituting Eq.~\eqref{eq14} into the Navier-Stokes equations and collecting terms order by order in $\epsilon$, we obtain a hierarchy of balances. Assuming quasi-stationarity, homogeneity, and isotropy, the order $O(\epsilon^0)$ balance recovers the standard ``3/2" law for the longitudinal increments of the background field,
\be
S^b_3(\ell)\equiv\mean{(\delta_\ell u_b)^3}=\frac{3}{2}\varepsilon_{b} \ell \,,
\label{eq15}
\ee
where $\varepsilon_{b} \simeq 0.74 \varepsilon$ is the inverse flux associated with the background field (see Fig.~\ref{fig2}).

At the next order, $O(\epsilon)$, we obtain an equation for the two-point vortex-background correlation $C_{bv}(\ell) =\langle \mathbf u_v(\mathbf x)\cdot \mathbf u_b(\mathbf{x}+\mathbf{\bel}) \rangle$. Under the same assumptions, this balance gives (for a detailed derivation see Appendix~\ref{app:two_point_balances})
\be
S^{bv}_3(\ell)\equiv 
\mean{\delta_\ell u_v\lr{\delta_\ell u_b}^2}
=\frac{1}{2}\varepsilon_{bv} \ell \ .\
\label{eq16}
\ee
The mixed structure function in Eq.~\eqref{eq16} represents the vortex-background contribution to the energy-transfer balance. The associated flux $\varepsilon_{bv}$ is the difference between the total flux and the background contribution and is measured from Fig.~\ref{fig2} to be
$\varepsilon_{bv} \simeq 0.25 \varepsilon$.

We test the above prediction by computing the different contributions to the third-order longitudinal structure function decomposed into background and vortex fields,
\be
S_3(\ell) \equiv \mean{(\delta_\ell u)^3}=
\mean{(\delta_\ell u_b)^3}+
3\mean{\delta_\ell u_v(\delta_\ell u_b)^2}+
3\mean{\delta_\ell u_b(\delta_\ell u_v)^2}+
\mean{(\delta_\ell u_v)^3}\ .\
\label{eq17}
\ee
The first two terms correspond to Eqs.~\eqref{eq15} and \eqref{eq16}, respectively, while the other two are of higher order in $\epsilon$, thus justifying the decomposition of Eq.~\eqref{eq14}.

\begin{figure}[htbp]
\centering\includegraphics[width=0.6\linewidth]{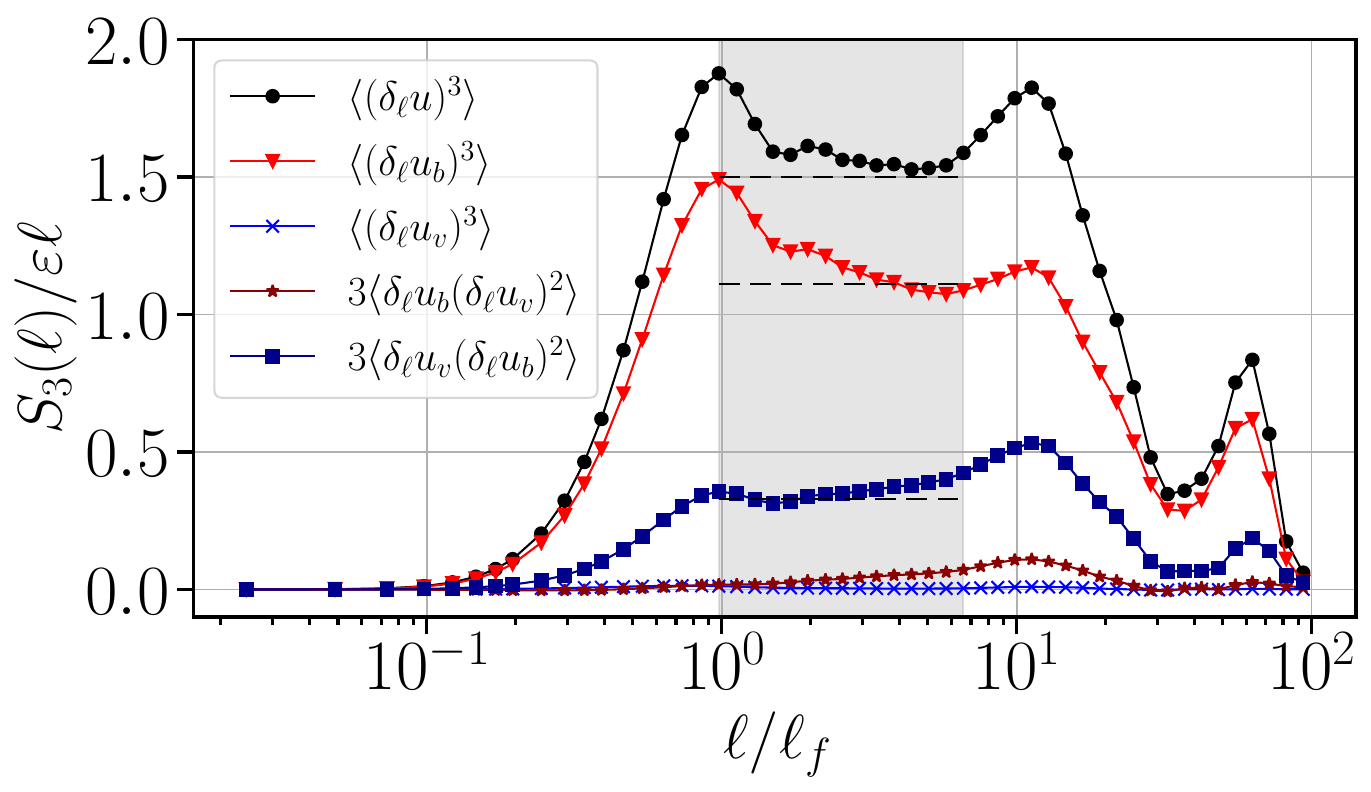}
\caption{Contributions to the third-order structure function (see legend), averaged over the time window $t/\tau_L\in[0.334,0.348]$. The plateau is expected to extend up to $L_I/\ell_f\approx 6.3$ (gray area). Dashed lines show the predicted plateau values.}
\label{fig6}
\end{figure}

Figure~\ref{fig6} shows the different contributions to the total third-order structure function computed at the final stage of the simulations. As expected, the relevant contributions are the first two terms in Eq.~\eqref{eq17}, while the term $\mean{\delta_\ell u_b(\delta_\ell u_v)^2}$ is very small. The plateau observed in $S_3^b(\ell)$ agrees with Eq.~\eqref{eq15}, with coefficient $\frac{3}{2} \varepsilon_b \simeq 1.125 \varepsilon$. Similarly, the plateau of $3 S_3^{bv}(\ell)$ agrees with Eq.~\eqref{eq16}, giving $\frac{3}{2}\varepsilon_{bv} \simeq 0.375 \varepsilon$. The linear scaling predicted by Eq.~\eqref{eq16} and verified by the simulations can be interpreted as a constant flux associated with the exchange between the background and vortex fields. This exchange produces a constant growth rate of the correlation $C_{bv}$, as shown in Fig.~\ref{fig2}.


\section{Conclusion and perspectives}

In this work, we studied the effect of small-scale vortices on the statistics of the inverse cascade in high-resolution numerical simulations of the 2D Navier-Stokes equations. We decomposed the vorticity and velocity fields into vortex and background components using the swirling-strength criterion.

The decomposition shows that vortices carry only a small fraction of the total energy and enstrophy, but have a measurable effect on the energy spectrum. In particular, the total spectrum does not display a clean Kolmogorov scaling, while the spectrum of the background field is much closer to the dimensional prediction. 
The vortex population becomes slowly varying at late times, with a radius distribution peaked near the forcing scale.
The simultaneous growth of $\sigma_R$ shows that the distribution continues to broaden even after the number of vortices and the mean radius have almost saturated. At the same time, the vortex kinetic energy continues to increase, indicating that the late-time evolution is controlled not by single-vortex statistics alone, but also by collective organization and vortex-background correlations. A remarkable phenomenological conclusion of our analysis is that, although the 3/2 law is well satisfied by both the total velocity field and its smooth background component, only the latter follows the Batchelor-Kraichnan dimensional prediction $E(k)\propto k^{-5/3}$. The spectrum of the total velocity field, instead, is affected by the vortex structures.

The scale-by-scale analysis clarifies the dynamical role of these correlations. The vortex field itself has negligible spectral flux and therefore does not directly carry the inverse cascade. Instead, most of the inverse-cascade flux is carried by the background field, while the remaining contribution appears through vortex-background correlations. This picture is supported in physical space by a generalized ``3/2" law for the mixed third-order structure function. The predicted linear scaling is observed in the simulations and can be interpreted as a constant inertial-range flux associated with the growth of vortex-background correlations.

These results suggest that coherent structures can influence equal-time scaling properties without acting as the direct carriers of the cascade flux. The decomposition developed here can be applied to other two-dimensional turbulent systems with small-scale coherent structures, including flows with large-scale friction, quasi-geostrophic dynamics, and laboratory realizations of 2D turbulence. A natural next step is to follow individual vortices in time, quantifying merger events, growth rates, lifetimes, and exchanges of circulation and enstrophy with the background field. Combining this with spatial diagnostics, such as vortex-vortex correlations, clustering measures, or entropy-based indicators, could clarify how their interactions and spatial distribution relate to the vortex-background energy transfer observed in the present work.

\section*{Acknowledgement}

V. J. V. acknowledges the support by the Italian Ministry of University and Research (MUR) - Fondo Italiano per la Scienza (FIS2) - 2023 Call, project DeepFL, CUP: E53C24003760001. L.M. acknowledges partial support from  CNPq and FAPERJ through Grants \# 311012/2022-1 and \# E-26/200.457/2026.
We acknowledge CINECA for computing resources through the INFN-CINECA grant INFN26-FieldTurb.

\bibliographystyle{ieeetr}
\bibliography{biblio}

@article{mcwilliams1990vortices,
  author = {McWilliams, J. C.},
  title = {The vortices of two-dimensional turbulence},
  journal = {J. Fluid Mech.},
  volume = {219},
  pages = {361--385},
  year = {1990}
}

@article{burgess2022statistics,
  author = {Burgess, B. H.},
  title = {Statistics of clustered vortices in the inverse energy cascade of two-dimensional turbulence},
  journal = {Phys. Rev. Fluids},
  volume = {7},
  pages = {104612},
  year = {2022}
}

@article{boffetta2012two,
  author = {Boffetta, G. and Ecke, R. E.},
  title = {Two-dimensional turbulence},
  journal = {Annu. Rev. Fluid Mech.},
  volume = {44},
  pages = {427--451},
  year = {2012}
}

@article{bernard1999three,
  author = {Bernard, D.},
  title = {Three-point velocity correlation functions in two-dimensional forced turbulence},
  journal = {Phys. Rev. E},
  volume = {60},
  pages = {6184--6187},
  year = {1999}
}

@article{boffetta2000inverse,
  author = {Boffetta, G. and Celani, A. and Vergassola, M.},
  title = {Inverse energy cascade in two-dimensional turbulence: Deviations from {Gaussian} behavior},
  journal = {Phys. Rev. E},
  volume = {61},
  pages = {R29--R32},
  year = {2000}
}

@article{muller2025lack,
  author = {M{\"u}ller, N. P. and Krstulovic, G.},
  title = {Lack of self-similarity in transverse velocity increments and circulation statistics in two-dimensional turbulence},
  journal = {Phys. Rev. Fluids},
  volume = {10},
  pages = {L012601},
  year = {2025}
}

@article{fontane2013vortical,
  author = {Fontane, J. J. L. and Dritschel, D. G. and Scott, R. K.},
  title = {Vortical control of forced two-dimensional turbulence},
  journal = {Phys. Fluids},
  volume = {25},
  pages = {015101},
  year = {2013}
}

@article{burgess2015kraichnan,
  author = {Burgess, B. H. and Scott, R. K. and Shepherd, T. G.},
  title = {{Kraichnan--Leith--Batchelor} similarity theory and two-dimensional inverse cascades},
  journal = {J. Fluid Mech.},
  volume = {767},
  pages = {467--496},
  year = {2015}
}

@article{smith1993bose,
  author = {Smith, L. M. and Yakhot, V.},
  title = {Bose condensation and small-scale structure generation in a random force driven {2D} turbulence},
  journal = {Phys. Rev. Lett.},
  volume = {71},
  pages = {352--355},
  year = {1993}
}

@article{chertkov2007dynamics,
  title={Dynamics of energy condensation in two-dimensional turbulence},
  author={Chertkov, M and Connaughton, C and Kolokolov, I and Lebedev, V},
  journal={Phys. Rev. Lett.},
  volume={99},
  number={8},
  pages={084501},
  year={2007},
  publisher={APS}
}

@article{frishman2018turbulence,
  author = {Frishman, A. and Herbert, C.},
  title = {Turbulence statistics in a two-dimensional vortex condensate},
  journal = {Phys. Rev. Lett.},
  volume = {120},
  pages = {204505},
  year = {2018}
}

@article{laurie2014universal,
  author = {Laurie, J. and Boffetta, G. and Falkovich, G. and Kolokolov, I. and Lebedev, V.},
  title = {Universal profile of the vortex condensate in two-dimensional turbulence},
  journal = {Phys. Rev. Lett.},
  volume = {113},
  pages = {254503},
  year = {2014}
}

@article{xu2024fluctuation,
  title={Fluctuation-induced transitions in anisotropic two-dimensional turbulence},
  author={Xu, Lichuan and van Kan, Adrian and Liu, Chang and Knobloch, Edgar},
  journal={Phys. Rev. Fluids},
  volume={9},
  number={6},
  pages={064605},
  year={2024},
  publisher={APS}
}

@article{lindborg1999can,
  author = {Lindborg, E.},
  title = {Can the atmospheric kinetic energy spectrum be explained by two-dimensional turbulence?},
  journal = {J. Fluid Mech.},
  volume = {388},
  pages = {259--288},
  year = {1999}
}

@article{kraichnan1967inertial,
  author = {Kraichnan, R. H.},
  title = {Inertial ranges in two-dimensional turbulence},
  journal = {Phys. Fluids},
  volume = {10},
  pages = {1417--1423},
  year = {1967}
}

@article{smith1994finite,
  author = {Smith, L. M. and Yakhot, V.},
  title = {Finite-size effects in forced two-dimensional turbulence},
  journal = {J. Fluid Mech.},
  volume = {274},
  pages = {115--138},
  year = {1994}
}

@article{paret1997experimental,
  author = {Paret, J. and Tabeling, P.},
  title = {Experimental observation of the two-dimensional inverse energy cascade},
  journal = {Phys. Rev. Lett.},
  volume = {79},
  pages = {4162--4165},
  year = {1997}
}

@article{paret1998intermittency,
  author = {Paret, J. and Tabeling, P.},
  title = {Intermittency in the two-dimensional inverse cascade of energy: Experimental observations},
  journal = {Phys. Fluids},
  volume = {10},
  pages = {3126--3136},
  year = {1998}
}

@article{zhou1999mechanisms,
  author = {Zhou, J. and Adrian, R. J. and Balachandar, S. and Kendall, T. M.},
  title = {Mechanisms for generating coherent packets of hairpin vortices in channel flow},
  journal = {J. Fluid Mech.},
  volume = {387},
  pages = {353--396},
  year = {1999}
}

@article{valadao2024spectrum,
  author = {Valad{\~a}o, V. J. and Boffetta, G. and De Lillo, F. and Musacchio, S. and Crialesi-Esposito, M.},
  title = {Spectrum correction in {Ekman-Navier-Stokes} turbulence},
  journal = {J. Turbul.},
  volume = {26},
  pages = {143--152},
  year = {2025}
}

@article{burgess2017scaling,
  title={Scaling theory for vortices in the two-dimensional inverse energy cascade},
  author={Burgess, B Helen and Scott, Richard K},
  journal={J. Fluid Mech.},
  volume={811},
  pages={742--756},
  year={2017},
  publisher={Cambridge University Press}
}

@article{farge1999non,
  author = {Farge, M. and Schneider, K. and Kevlahan, N.},
  title = {Non-{Gaussianity} and coherent vortex simulation for two-dimensional turbulence using an adaptive orthogonal wavelet basis},
  journal = {Phys. Fluids},
  volume = {11},
  pages = {2187--2201},
  year = {1999}
}


\clearpage 

\appendix

\section{Vortex-identification methods and threshold dependence}
\label{app:threshold_dependency}

To evaluate how our results depend on the vortex-identification procedure, we compare the vortex field obtained with the swirling-strength criterion used throughout the paper with two alternative decompositions. The first is Coherent Vorticity Extraction (CVE), based on a discrete wavelet decomposition \cite{farge1999non}. The second follows the area-based filtering strategy used in the vortex-identification procedure of Burgess and Scott \cite{burgess2017scaling}, in which structures below a prescribed forcing-scale-dependent area are discarded.

\begin{figure}[htbp]
\centering\includegraphics[width=\linewidth]{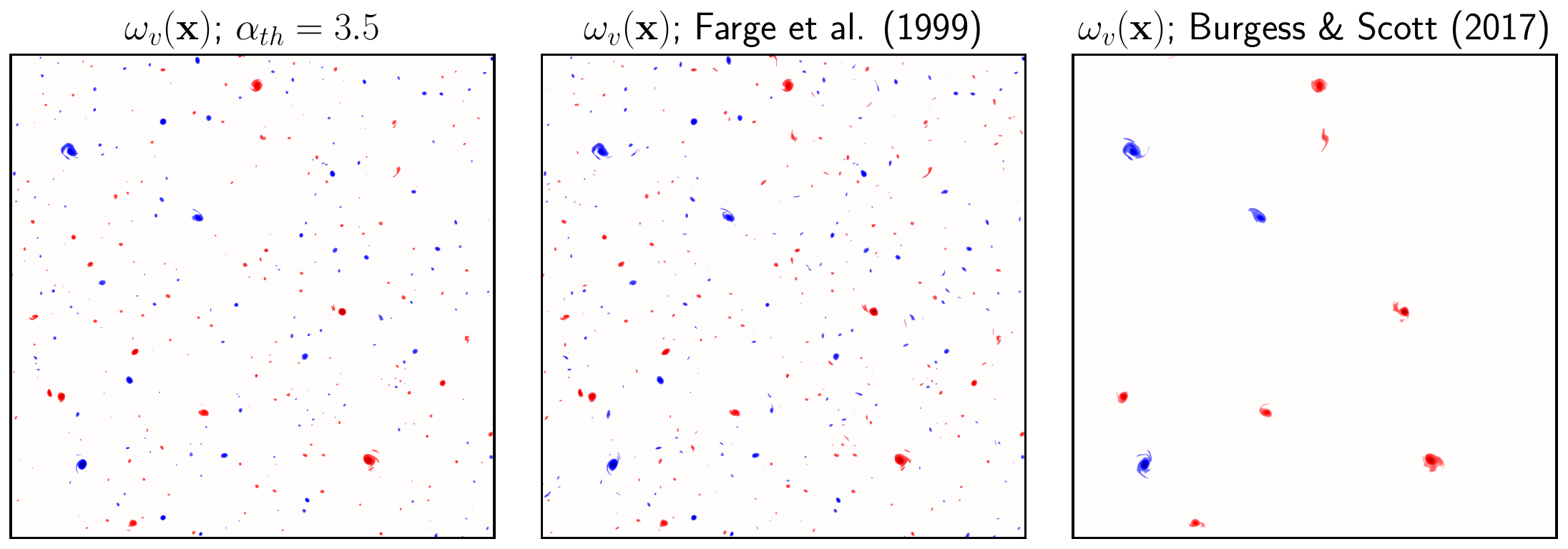}
\caption{Vortex components of the vorticity field $\omega_v$ at the final time for different decomposition methods. The figure shows a region of size $(L/4)^2$ of the periodic box.}
\label{figa1}
\end{figure}

Figure~\ref{figa1} shows that the swirling-strength and CVE decompositions identify a broadly similar population of compact vortical structures. In the CVE implementation, the vorticity field is decomposed into wavelet coefficients $c_\lambda$, which are separated into coherent and incoherent contributions through an iteratively determined threshold $T$. Starting from $T=q\omega_{\rm rms}$, the variance is recomputed using only the coefficients satisfying $|c_\lambda|\leq T$, and the threshold is updated until convergence. The coherent vortex field $\omega_v$ is then reconstructed from the coefficients with $|c_\lambda|>T$, while the remaining coefficients define the background field $\omega_b$. We use $q=5.95$, consistent with the universal-threshold prescription commonly employed in CVE, and the \texttt{sym6} wavelet basis with periodic boundary conditions. Since the coherent and incoherent fields are reconstructed from disjoint subsets of an orthogonal wavelet representation, one expects $\mean{\omega_v\omega_b}\simeq0$; in our simulations, the residual cross-correlation remains of order $10^{-8}\omega_{\rm rms}^2$.

The Burgess-type identification, instead, produces a much sparser vortex population because the area cutoff removes many forcing-scale structures. This makes it useful for isolating the largest coherent vortices, but less appropriate for quantifying the full population of vortices generated near the forcing scale in the present simulations. As a result, this method changes the measured vortex-size distribution significantly with respect to the swirling-strength and CVE decompositions. This methodological difference is one of the reasons why a direct comparison of vortex-size distributions with \cite{burgess2022statistics} is delicate, as discussed in the main text.

\begin{figure}[htbp]
\centering\includegraphics[width=0.49\linewidth]{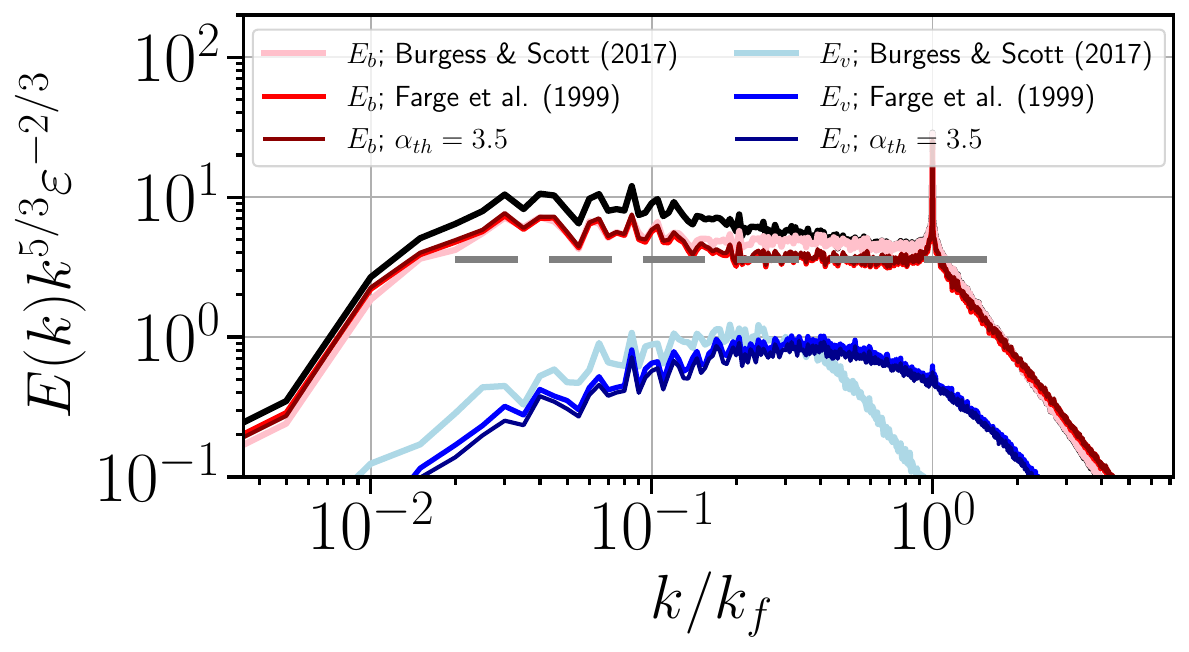}
\centering\includegraphics[width=0.47\linewidth]{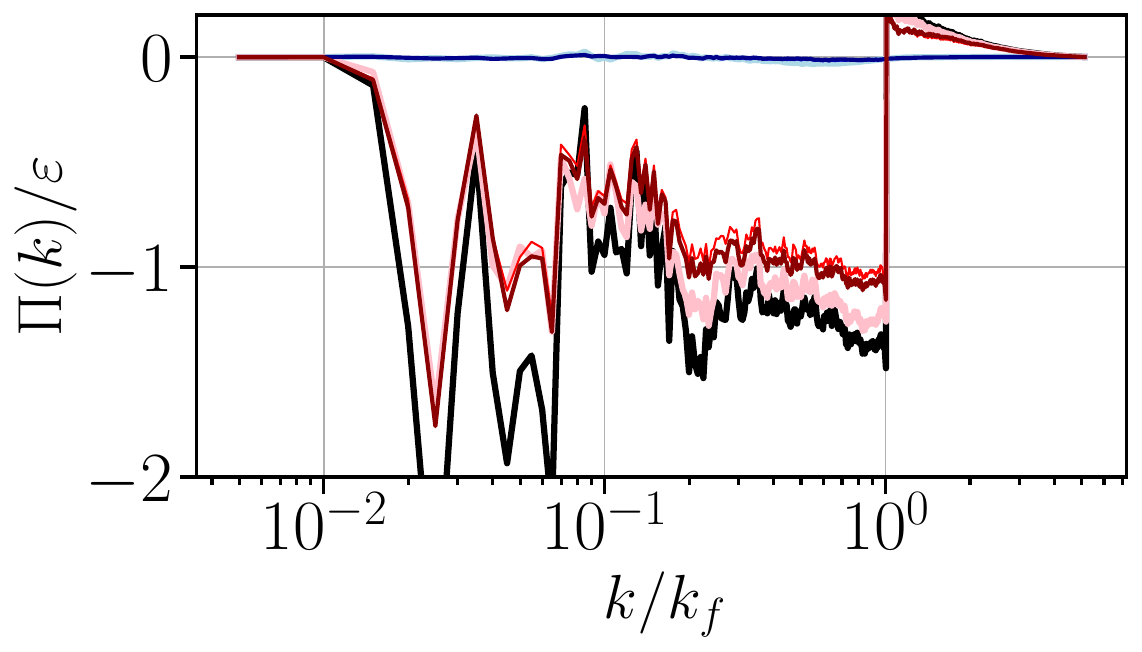}
\caption{Left: compensated energy spectra; right: nondimensionalized energy fluxes for the final snapshot of the simulation. The total field is shown in black. Blue tones denote vortex contributions and red tones denote background contributions. Light colors correspond to the Burgess-type area-filtering method, intermediate colors to CVE, and dark colors to the swirling-strength method used throughout the paper.
}
\label{figa2}
\end{figure}

The scale-by-scale diagnostics in Fig.~\ref{figa2} show that the Burgess-type area filtering also changes the compensated spectra more strongly than CVE. Since the small-scale vortices retained by the swirling-strength and CVE decompositions are instead included in the background by the Burgess-type filter, the amplitude of the compensated background spectrum changes. Equivalently, a different effective Kolmogorov constant appears for the background field. Nevertheless, the main conclusion remains unchanged: in all decompositions the vortex contribution carries a negligible flux, while the inverse-cascade flux is carried predominantly by the background field, with vortex-background correlations providing the remaining exchange term.

We therefore use the swirling-strength method because it is local, simple to implement, and directly tied to the velocity-gradient topology. Its only free parameter is the threshold $\alpha$ in Eq.~(7). We determine this value from the fact that, once the coherent vortical contribution is removed, the one-point statistics of the background vorticity should be close to those of a symmetric Gaussian field.
We therefore define the cost function
\be
S(\alpha)=\lr{\frac{\mean{\omega_b^3}}{\mean{\omega_b^2}^{3/2}}}^{2}+
\lr{\frac{\mean{\omega_b^4}}{\mean{\omega_b^2}^2}-3}^2 \ ,\
\label{eqa1}
\ee
which measures the squared skewness and squared excess flatness of the background vorticity.
\begin{figure}[htbp]
\centering
\includegraphics[width=0.5\linewidth]{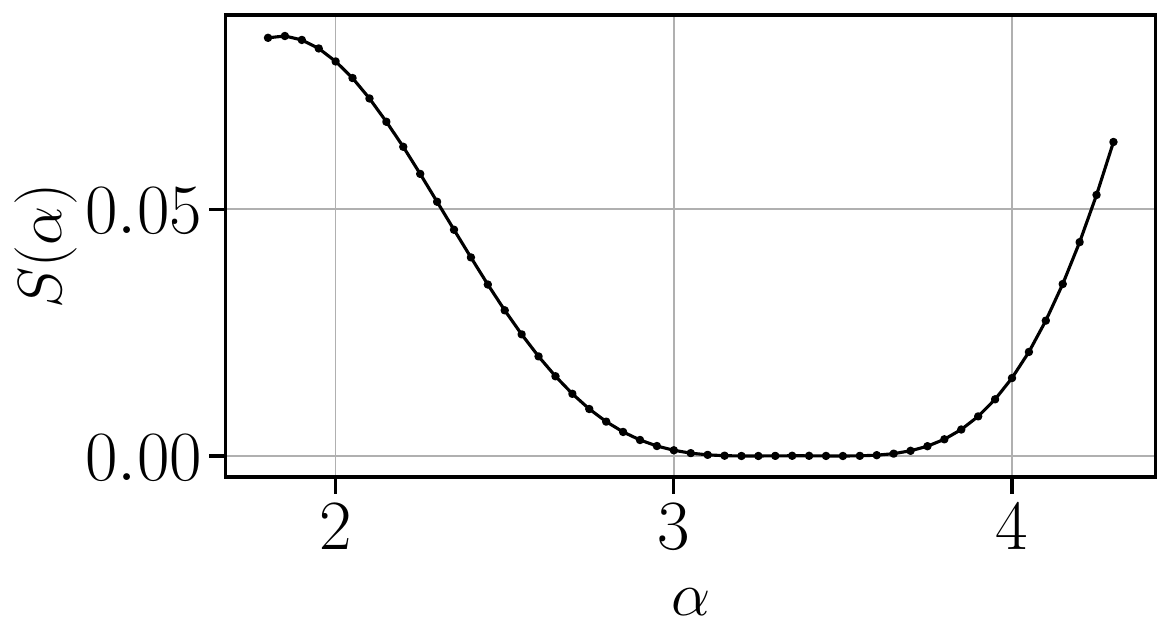}
\caption{Cost function as a function of the threshold in the swirling-strength method, computed for one vorticity field $\omega$ at the final simulation time.}
\label{figa3}
\end{figure}

Figure~\ref{figa3} shows that $S\approx0$ in a range that spans $\alpha\in[3.1,3.6]$, with an actual minimum between $\alpha=3.45$ and $\alpha=3.5$. We then choose $\alpha=3.5$ throughout the paper.

\section{Karman--Howarth balances for the background and vortical fields}
\label{app:two_point_balances}

In this Appendix, we provide a formal derivation of the two-point balances associated with the asymptotic decomposition
\be
\mathbf u = \mathbf u_b + \epsilon \mathbf u_v,
\qquad
p = p_b + \epsilon p_v,
\qquad
0<\epsilon\ll 1,
\label{eqb1}
\ee
with both fields incompressible,
\be
\nabla\cdot \mathbf u_b = 0 \ ,
\qquad
\nabla\cdot \mathbf u_v = 0 \ .
\label{eqb2}
\ee
The motivation for this expansion is that the vortical contribution carries only a small fraction of the kinetic energy, while the background field contains the dominant part of the inverse-cascade dynamics. Throughout this Appendix, we neglect forcing and viscous terms because our purpose is only to isolate the inertial-range transfer terms. Starting from the incompressible Euler equations
\be
\partial_t \mathbf u + (\mathbf u\cdot\nabla) \mathbf u = - \nabla p,
\label{eqb3}
\ee
and substituting \eqref{eqb1}, one obtains
\be
\partial_t \mathbf u_b
+\epsilon \partial_t \mathbf u_v
+ (\mathbf u_b\cdot\nabla) \mathbf u_b
+\epsilon\Big(
(\mathbf u_b\cdot\nabla) \mathbf u_v
+(\mathbf u_v\cdot\nabla) \mathbf u_b
\Big)
+\epsilon^2 (\mathbf u_v\cdot\nabla) \mathbf u_v
=
-\nabla p_b - \epsilon \nabla p_v.
\label{eqb4}
\ee
Collecting terms order by order yields
\be
\partial_t \mathbf u_b + 
(\mathbf u_b\cdot\nabla) \mathbf u_b 
= -\nabla p_b,
\label{eqb5}
\ee
and
\be
\partial_t \mathbf u_v
+(\mathbf u_b\cdot\nabla) \mathbf u_v
+(\mathbf u_v\cdot\nabla) \mathbf u_b
=
-\nabla p_v.
\label{eqb6}
\ee
Equation \eqref{eqb5} is the standard incompressible Euler equation for the background field. Equation \eqref{eqb6} is the linearized dynamics of the vortical field around the background flow. The quadratic self-interaction $(\mathbf u_v\cdot\nabla) \mathbf u_v$ only enters at order $O(\epsilon^2)$ in the asymptotic expansion.

The background field obeys the standard inviscid incompressible dynamics in Eq.~\eqref{eqb5}. Therefore, the derivation of its two-point balance is canonical. The quantity $\mean{|\mathbf u_b|^2}$ is conserved, and its two-point correlation, under the assumptions of statistical homogeneity and isotropy, obeys the Karman--Howarth--Monin equation
\be
\partial_t C_{bb}
-\frac{1}{4}\nabla_{\ell}\cdot
\mean{
\delta_\ell \mathbf u_b\, |\delta_\ell \mathbf u_b|^2
}=0\ ,\
\label{eqb7}
\ee
where $\delta_\ell \mathbf u_b=\mathbf  u_b(\mathbf{x}+\mathbf{\bel})-\mathbf u_b(\mathbf{x})$ and
\be
C_{bb}(\ell,t) =\frac{1}{2}\mean{
 \mathbf u_b(\mathbf x)\cdot \mathbf u_b(\mathbf{x}+\mathbf{\bel})}\ .\
\label{eqb8}
\ee
Assuming quasi-stationarity and a constant inverse-cascade flux $\varepsilon_{b}$, one gets
\be
\nabla_\ell\cdot
\mean{
\delta_\ell \mathbf u_b\, |\delta_\ell \mathbf u_b|^2
}=4\varepsilon_{b}\ ,\
\label{eqb9}
\ee
which in two dimensions gives the standard ``3/2" law for longitudinal increments,
\be
S^b_3(\ell)\equiv\mean{(\delta_\ell \mathbf u_b\cdot\hat{\bel})^3}=\frac{3}{2}\varepsilon_{b}\ell \ .\
\label{eqb10}
\ee

The vortex self-energy $E_v$ is not conserved because its balance contains the production term
\[
P_v=\mean{\mathbf u_v\cdot [(\mathbf u_v\cdot \nabla )\mathbf u_b]}.
\]
This production is compensated by the missing $\mathcal{O}(\epsilon^2)$ term in the total energy balance. It does not, however, affect the balance of the correlation $\mean{\mathbf u_b\cdot\mathbf u_v}$ at order $\mathcal{O}(\epsilon)$. Defining the two-point mixed correlation
\be
C_{bv}(\ell,t) =\mean{
 \mathbf u_v(\mathbf x)\cdot \mathbf u_b(\mathbf{x}+\mathbf{\bel})}\ ,\
\label{eqb11}
\ee
one can use \eqref{eqb5}, \eqref{eqb6}, and \eqref{eqb11} to derive the evolution of $C_{bv}$:
\be
\partial_t C_{bv}
-\frac{1}{4}\nabla_{\ell}\cdot\bigg(
\mean{\delta_\ell \mathbf u_v\, |\delta_\ell \mathbf u_b|^2}+
2\mean{\delta_\ell \mathbf u_b (\delta_\ell \mathbf u_v\cdot\delta_\ell \mathbf u_b)}
\bigg)=0\ .\
\label{eqb12}
\ee
The two flux terms in Eq.~\eqref{eqb12} have distinct physical interpretations. The first contribution originates from $(\mathbf u_v\cdot\nabla) \mathbf u_b$ in Eq.~\eqref{eqb6} and represents the advection of background energy by the vortical field. The second contribution combines $(\mathbf u_b\cdot\nabla) \mathbf u_b$ and $(\mathbf u_b\cdot\nabla) \mathbf u_v$ in the correlation balance, producing a flux term of the form $\nabla\cdot[(\mathbf u_b\cdot\mathbf u_v) \mathbf u_b]$. It therefore measures the transport, by the background field, of local alignment between vortex and background velocities.

Assuming quasi-stationarity again and denoting the corresponding mixed inverse-cascade flux by $\varepsilon_{bv}$, one gets
\be
F^{L}(\ell)=\mean{\delta_\ell u^L_v |\delta_\ell \mathbf u_b|^2}+
2\mean{\delta_\ell u^L_b (\delta_\ell \mathbf u_v\cdot\delta_\ell \mathbf u_b)}=
2\varepsilon_{bv}\ell \ ,\
\label{eqb13}
\ee
where the increments are decomposed into longitudinal and transverse components, $\delta\mathbf  u=\delta_\ell u^L \hat{\bel}+\delta_\ell u^T \hat{\mathbf t}$. Further expansion of the left-hand side gives three different contributions to the correlation flux:
\be
F^{L}(\ell)=3\mean{\delta_\ell u^L_v\lr{\delta_\ell u^L_b}^2}+
2\mean{\delta_\ell u^T_v \delta_\ell u^T_b \delta_\ell u^L_b}+
\mean{\delta_\ell u^L_v\lr{\delta_\ell u^T_b}^2}=2\varepsilon_{bv}\ell \ .\
\label{eqb14}
\ee
To relate this flux relation to a longitudinal structure function, we introduce the symmetrized mixed third-order tensor
\be
M_{ijk}(\ell)
=
\left\langle
\delta_\ell u_{v,i}\delta_\ell u_{b,j}\delta_\ell u_{b,k}
+\delta_\ell u_{b,i}\delta_\ell u_{v,j}\delta_\ell u_{b,k}
+\delta_\ell u_{b,i}\delta_\ell u_{b,j}\delta_\ell u_{v,k}
\right\rangle .
\ee
With this definition, the mixed flux in Eq.~(B13) is
\be
F^{L}(\ell)=\hat{\ell}_i M_{ijj}(\ell)=2\varepsilon_{bv}\ell,
\ee
while its fully longitudinal component is
\be
M_{LLL}(\ell)=3\left\langle
\delta u_v^L(\delta u_b^L)^2
\right\rangle .
\ee
For an isotropic incompressible third-order tensor in d-dimensions, one has
\be
F^{L}(\ell)=\frac{d+2}{3}M_{LLL}(\ell)
=\frac{4}{3}M_{LLL}(\ell).
\ee
Therefore Eq.~(B13) gives
\be
S^{bv}_3(\ell)\equiv 
\mean{\delta_\ell u^L_v\lr{\delta_\ell u^L_b}^2}
=\frac{1}{2}\varepsilon_{bv}\ell \ .\
\label{eqb15}
\ee
In the main text, the energy growth rates give $\varepsilon_{b}\approx3\varepsilon/4$ and $\varepsilon_{bv}\approx\varepsilon/4$. Therefore, the background-background contribution to the third-order structure function has a plateau of approximately $1.125\varepsilon$, while the mixed contribution has a plateau of approximately $0.375\varepsilon$.

\begin{figure}[htbp]
\centering
\includegraphics[width=0.6\linewidth]{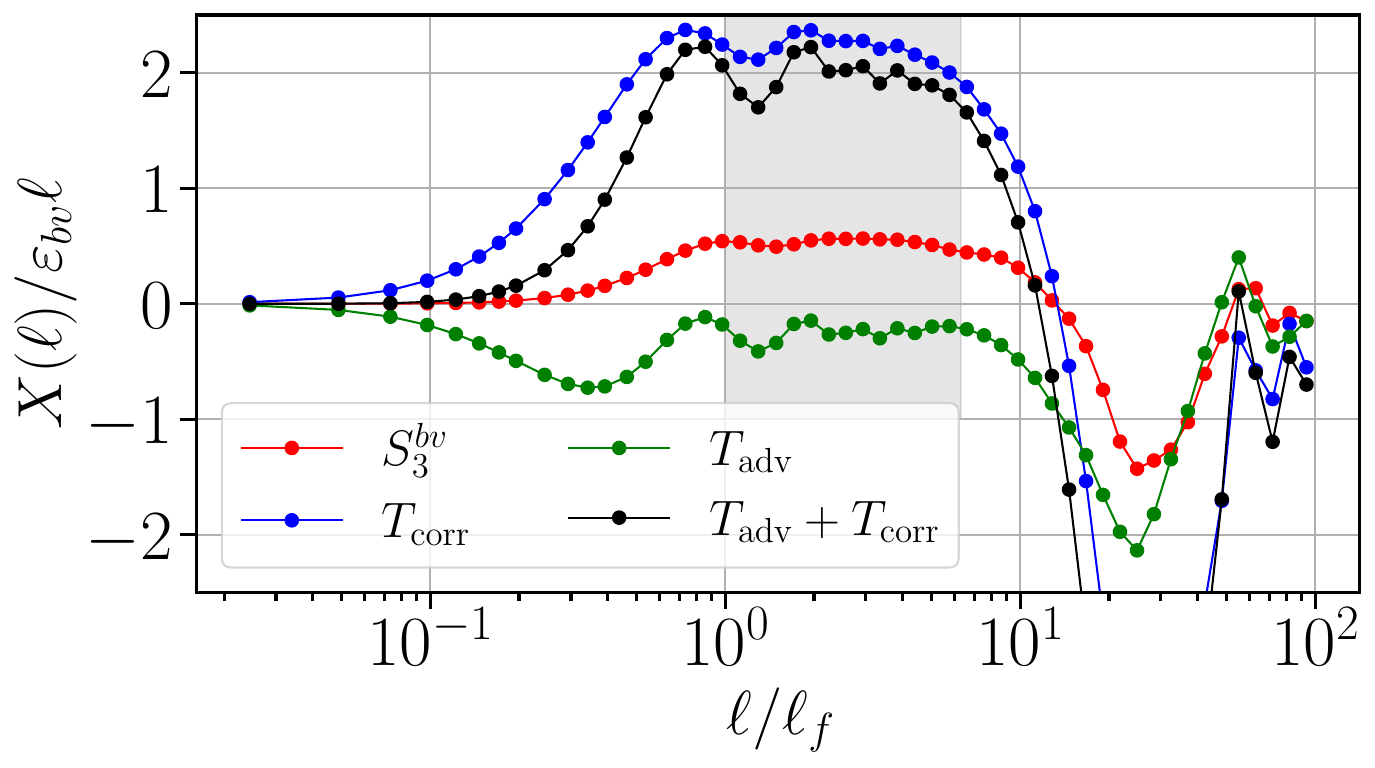}
\caption{Contributions to equations \eqref{eqb13} and \eqref{eqb15} (see legend). Data were averaged over the time window $t/\tau_L\in[0.334,0.348]$. The plateaus are expected to extend up to $L_I/\ell_f\approx 6.3$ (gray area).}\label{figb1}
\end{figure}

To illustrate the content of Eq.~\eqref{eqb13}, Fig.~\ref{figb1} shows the two contributions to the mixed correlation flux, computed from one realization and averaged over 16 nearby snapshots, representing the same time interval as in Fig.~\ref{fig6}. All contributions show a scaling compatible with the dimensional $\propto \vare_{bv}\ell$ with different proportionality constants. The measured $S_3^{bv}/\vare_{bv}\ell=0.53\pm0.03$ is compatible with the expected $1/2$ value. The term
\be
T_{\rm adv}(\ell)=
\mean{\delta_\ell u_v^L |\delta_\ell \mathbf u_b|^2}
\ee
is associated with the advection of background-energy fluctuations by the vortex velocity field. It is negative over most of the inertial range ($T_{\rm adv}/\vare_{bv}\ell=-0.25\pm0.07$), suggesting a downscale contribution to the mixed transfer. 
By contrast, the correlation-transport term
\be
T_{\rm corr}(\ell)=
2\mean{\delta_\ell u_b^L
(\delta_\ell \mathbf u_v\cdot\delta_\ell \mathbf u_b)}
\ee
is positive and dominates the balance ($T_{\rm corr}/\vare_{bv}\ell=2.21\pm0.10$). This term can be interpreted as the inertial transport, by the background field, of local vortex-background alignment. Their sum is positive over the inertial range and is close to the prediction of \eqref{eqb13}
\be
T_\text{adv} +T_{\rm corr}(\ell)
\simeq A\varepsilon_{bv}\ell ,
\ee
with $A=1.95\pm0.14$, supporting the interpretation that the background flow not only carries most of the inverse energy flux, but also organizes its correlations with the vortical field across scales.

\end{document}